# Monolayer RhB$_4$: half-auxeticity and almost ideal spin-orbit Dirac point semimetal

Zhen Gao,[1] Qianqian Wang,[2] Weikang Wu,[2,3,*] Zhixue Tian,[1] Ying Liu,[1] Fengxian Ma,[1,*] Yalong Jiao,[1,4] Shengyuan A. Yang[2]

[1] College of Physics, Hebei Key Laboratory of Photophysics Research and Application, Hebei Normal University, Shijiazhuang 050024, China

[2] Research Laboratory for Quantum Materials, Singapore University of Technology and Design, Singapore 487372, Singapore

[3] Division of Physics and Applied Physics, School of Physical and Mathematical Sciences, Nanyang Technological University, Singapore 637371, Singapore

[4] Faculty for Chemistry and Food Chemistry, TU Dresden, Bergstraße 66c, Dresden 01069, Germany

*E-mail: weikang.wu@ntu.edu.sg, fengxianma@hebtu.edu.cn

**ABSTRACT**
Structural-property relationship, the connection between materials' structures and their properties, is central to the materials research. Especially at reduced dimensions, novel structural motifs often generate unique physical properties. Motivated by a recent work reporting a novel half auxetic effect in monolayer PdB$_4$ with a hypercoordinated structure, here, we extensively explore similar 2D transition metal boride structures $M$B$_4$ with $M$ covering 3$d$ and 4$d$ elements. Our investigation screens out one stable candidate, the monolayer RhB$_4$. We find that monolayer RhB$_4$ also shows half auxeticity, i.e., the material always expands in a lateral in-plane direction in response to an applied strain in the other direction, regardless of whether the strain is positive or negative. We show that this special mechanical character is intimately tied to the hypercoordinated structure with the $M$©B$_8$ structural motif. Furthermore, regarding electronic properties, monolayer RhB$_4$ is found to be the first example of an almost ideal 2D spin-orbit Dirac point semimetal. The low-energy band structure is clean, with a pair of fourfold degenerate Dirac points robust under spin-orbit coupling located close to the Fermi level. These Dirac points are *enforced* by the nonsymmorphic space group symmetry which is also determined by the lattice structure. Our work deepens the fundamental understanding of structural-property relationship in reduced dimensions. The half auxeticity and the spin-orbit Dirac points will make monolayer RhB$_4$ a promising platform for nanomechanics and nanoelectronics applications.

## I. INTRODUCTION

The connection between structures and their properties, the structure-property relationship, is fundamental to multiple fields, ranging from mathematics, chemistry, physics, engineering product design, to social sciences. It is particularly appreciated in the study of solid materials [1,2], where the structure can be precisely imaged and its impacts on material properties can be analyzed in a quantitative way. As prominent examples, graphite and diamond are both allotropes of carbon [3], but their different structures generate dramatically different properties. On the other hand, materials possessing similar structures may share similar properties. And special structural motifs may lead to certain unique properties.

Like carbon, boron is a light element that also features rich polymorphism due to its capability to form various types of hybridized bonding [4-6]. In recent years, there is considerable effort in achieving two-dimensional (2D) structures of boron [7-22]. Due to the electron deficiency character, the stability of such structures would require electron doping either from metallic substrates or by forming 2D compounds with metal elements. Interestingly, it was found that when combined with certain transition metal elements, 2D boride materials may possess special hypercoordinated structures. For example, this was predicted in monolayer FeB$_6$ [8], where each Fe is bonded to 6 or 8 nearby B sites forming a Fe©B$_x$ ($x = 6, 8$) wheel-like structure. Very recently, hypercoordinated structural motif was also reported in monolayer PdB$_4$ [23], which features periodic Pd©B$_8$ wheel structures. Importantly, due to the larger size of Pd, the wheel is "corrugated", and this special structure leads to an exotic half auxetic behavior, namely, the material always expands in one direction, regardless of whether it is compressed or pulled in the orthogonal direction.

Two questions remain after the discovery in Ref. 23. First, is the connection between the corrugated wheel structure and the half auxeticity generic? In other words, can we find other examples that also demonstrate this connection? Second, can the structural characters also manifest in other properties, such as electronic structures?

To address these questions, in this work, we extensively explore the possibility of forming 2D transition metal boride structures with the $M$©B$_8$ structural motifs, where $M$ covers the 3$d$ and 4$d$ elements. Based on the density functional theory (DFT) calculations, we screen out one stable candidate, the monolayer RhB$_4$. We show that the material also exhibits half auxeticity, for which the connection to the Rh©B$_8$ structural motif is carefully analyzed and exposed. Furthermore, the hypercoordinated structure with corrugation endows the lattice with nonsymmorphic symmetries, i.e., crystal symmetries involving fractional lattice translations. We show that the nonsymmorphic symmetries enforce pairs of fourfold degenerate Dirac points in the electronic band structure, which are robust under spin-orbit coupling (SOC). In monolayer $RhB_4$, two such Dirac points are located very close to the Fermi level, making the material the first example of an almost ideal 2D spin-orbit Dirac point semimetal. Our work not only reveals fascinating consequences of the hypercoordinated structural motifs in 2D boron compounds, but also predicts a promising platform for studying half auxeticity and spin-orbit Dirac points.

## II. COMPUTATIONAL DETAILS

The Vienna ab initio simulation package (VASP) [24]was used for the DFT calculations and the projector augmented wave method (PAW) approach [25] was adopted to treat the ion-electron interactions. The electronic exchange-correlation functional was treated in the generalized gradient approximation (GGA) by using the one parametrized by Perdew, Burke, and Ernzerhoff (PBE) [26]. To deal with the correlation effects for the *4d* orbitals of Rh, the Hubbard *U* correction was used with the value of 3.0 eV [27]. We have tested *U* values from 0 to 4 eV and found no qualitative change to our results from the correlation effects (see Fig. S1 in Supplemental Material) [28]. And our key results were also confirmed by the calculation with hybrid functional approach (HSE06) (Fig. S2) [28]. The energy cutoff of the plane waves was set to 400 eV. The structures were fully relaxed until the force on each atom was less than 0.005 eV/Å. The energy convergence criterion in the self-consistent calculations was set to $10^{-5}$ eV. $6 \times 6 \times 1$ and $12 \times 12 \times 1$ Monkhorst−Pack *k*-point grids were used for geometry optimization and self-consistent calculation, respectively. A vacuum slab of at least 20 Å in z direction was adopted to avoid artificial interactions between the neighboring layers. The phonon dispersion was computed by using the Phonopy code [29] within the density functional perturbation theory (DFPT) [30]. Thermal stabilities were evaluated by using ab initio molecular dynamics (AIMD) simulations. The strain is defined as $\varepsilon = (a - a_0)/a_0$, where a is the lattice parameter in strained state and $a_0$ represents that for the strain-free state.

## III. RESULTS AND DISCUSSION

As mentioned, we have explored 2D *M*$B_4$ compounds with lattice structures similar to that of $PdB_4$. For *M* running over the 3*d* and 4*d* elements and considering non-magnetic compounds with the symmetry of $PdB_4$, we find that only $RhB_4$ can be stable in this 2D structure. The crystal structure of monolayer $RhB_4$ is shown in Fig. 1. One can see that the B atoms form a network composed of 4-membered and 8-membered rings, and each metal atom Rh is sitting at the center of an 8-membered B ring and is octacoordinated to B atoms, forming the Rh©$B_8$ wheel-like structural motifs. Importantly, the wheel is corrugated, namely, four B atoms are in the Rh atomic plane, whereas the other four B atoms in a wheel are located off the plane, as illustrated in Fig. 1. This corrugation plays a crucial role in our discussion below.

The crystal belongs to the *Cmma* space group (No. 67). The optimized lattice parameters are a = b =5.487 Å, and the thickness of layer is 1.38 Å, as indicated in Fig. 1c. The optimized average Rh-B and B-B bond lengths are 2.19Å and 1.7 Å, respectively. To characterize the bonding behavior, the electron localization function (ELF) was calculated (Fig. 1e). Basically, we observe that ELF is peaked between B-B atoms, whereas it is suppressed between B-Rh atoms, reflecting the strong covalent bonding between B atoms and the ionic bonding between B and Rh.

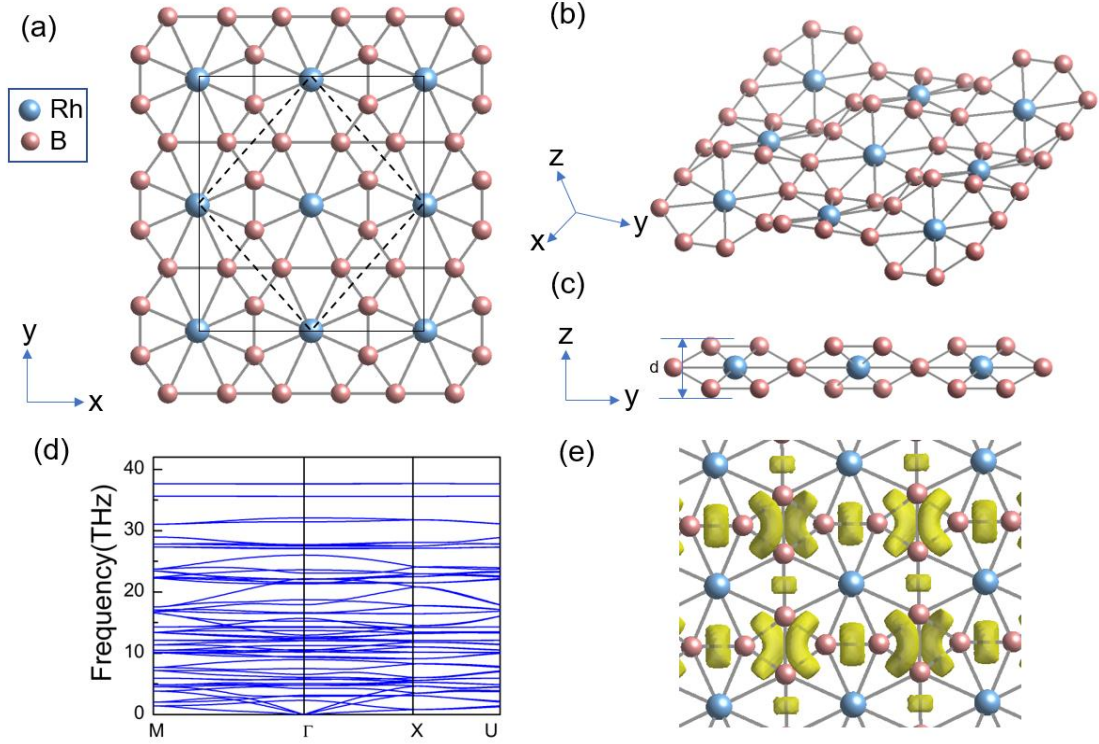

Fig. 1 (a-c) Atomic structure of the RhB$_4$ monolayer. The dashed line in (a) represents the primitive cell and the solid line marks a rectangular unit cell. (d) The phonon dispersion of the RhB$_4$ monolayer. (e) The electron localization function (ELF) with the isovalue of 0.75 eÅ$^{-3}$.

To test the structural stabilities of the monolayer, we first calculated the cohesive energy $E_{co}$ based on the definition:

$$E_{co} = [E(Rh_xB_y) - xE(Rh) - yE(B)]/(x+y)$$

where *x* and *y* indicate the number of corresponding atom in a unit cell, $E(Rh_xB_y)$ is the total energy of the RhB$_4$ monolayer, E(Rh) and E(B) are the energies of isolated Rh and B atoms, respectively. We find that $E_{co}$ for RhB$_4$ monolayer is -6.71 eV per atom, which is comparable with that of graphene (-7.85 eV per atom), and is lower than FeB$_2$ (-4.87 eV per atom), FeB$_3$ (-5.93 eV per atom) and FeB$_6$ (-5.56 to -5.79 eV per atom) [31-34], indicating its good stability.

According to the elastic stability criteria [35], stable 2D lattices should satisfy $C_{11}, C_{22}, C_{66} > 0$ and $C_{11} + C_{22} - 2C_{12} > 0$, where $C_{ij}$ are the elastic constants. For monolayer RhB$_4$, the calculated values of these constants are $C_{11} = 146.8$ N/m, $C_{22} = 129.9$ N/m and $C_{66} = 65.0$ N/m. One confirms that the criteria are satisfied, indicating the mechanical stability of the material.

The dynamically stability can be inferred from the phonon dispersions. The calculation result is shown in Fig. 1d. One can see that there is no imaginary mode in the entire Brillouin zone, so the RhB$_4$ monolayer is dynamically stable. We further confirmed the thermal stability of the material by performing AIMD simulations at 300 K and 500 K. As shown in Fig. S3 [28], the energy of monolayer RhB$_4$ fluctuates only in a small range (~0.1 eV/atom) and its structure can be well maintained through the simulation time.

In elasticity theory, 2D Young's moduli along x and y directions are defined by the elastic constants as $Y_x^{2D} = (C_{11}C_{22} - C_{12}C_{21})/C_{22}$ and $Y_y^{2D} = (C_{11}C_{22} - C_{12}C_{21})/C_{11}$, respectively. The evaluated $Y_x^{2D}$ and $Y_y^{2D}$ values are 146.8 and 129.9 GPa·nm for monolayer RhB$_4$, respectively. These values are about 39 - 44% of that in 2D MoS$_2$ (330 GPa·nm) [36], but larger than that of phosphorene (23.0 - 92.3 GPa·nm) [37], demonstrating it is a flexible material.

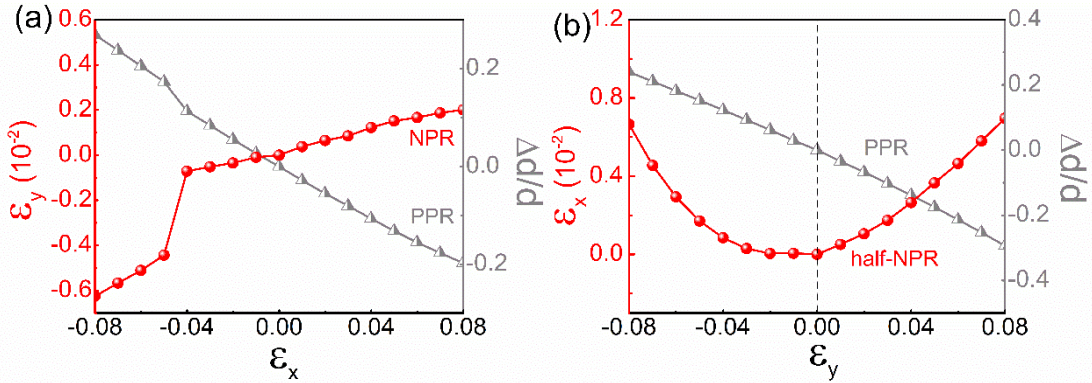

Fig. 2 Mechanical response of the RhB$_4$ monolayer under uniaxial strain along (a) x and (b) y directions. PPR and NPR represent positive or negative PR.

Unique structures usually lead to novel mechanical properties in 2D materials such as negative Poisson's ratio (PR) [38]. To further study the mechanical properties of monolayer RhB$_4$, uniaxial strains ranging from -8% to 8% are applied along one in-plane direction, and we look for the responsive strains along the other two perpendicular directions. The results are shown in Fig. 2. First, from the black curves in the two figures, one observes that the layer thickness decreases under both $\varepsilon_x$ and $\varepsilon_y$, indicating that the out-of-plane PR is positive. This is the usual behavior. Second, in Fig. 2a, under applied strain $\varepsilon_x$, the resultant $\varepsilon_y$ curve exhibits a positive slope, indicating a negative PR. In other words, the material expands (shrinks) along y when stretched (compressed) along x. This is the unusual auxetic behavior. (Note that there is an abrupt change in the curve at $\varepsilon_x \sim -5\%$, which is related to a structural phase transition.) Third and most interestingly, in Fig. 2b, under applied strain $\varepsilon_y$, the curve for the responsive strain $\varepsilon_x$ shows a concave shape, with positive values for both positive and negative $\varepsilon_y$. This is the most exotic half auxetic behavior, as reported in Ref. 23 for 2D PdB$_4$. Essentially, it means the sheet always expands along y, regardless of whether it is stretched or compressed along y.

The half auxeticity of PdB$_4$ was discussed in Ref. 23. Here, we will try to understand it from another intuitive perspective. We will show that the behavior is a consequence of their special and share structural motif, namely, the hypercoordinated M©B8 wheels. To visualize this feature more clearly, we plot an enlarged view of the Rh©B8 wheel in Fig. 3. One important point is that the wheel is corrugated. One can see that the Rh atom and B atoms labeled 1, 2, 5, and 6 are within a plane (the z=0 plane), while B 3, 4 and B 7, 8 are below and above this plane, respectively. This corrugation enables vertical motions of the atoms under an in-plane strain, which is critical for their peculiar mechanical property.

First, let's consider the negative PR when strain $\varepsilon_x$ is applied (see Fig. 2a). Under the tensile $\varepsilon_x$ strain (Fig. 3a), the four out-of-plane B atoms will naturally move towards the z=0 plane. This movement will push out B 1, 2, 5 and 6 along y, because of the strong covalent bonding between the B atoms such that the bond lengths between the B sites are kept more or less unchanged (~1.66 Å). Conversely, when the wheel is compressed along x (Fig. 3b), the four out-of-plane B will move further away from the z=0 plane, and the four in-plane B will then shift towards Rh. This explains the observed auxetic behavior for applied strains along x.

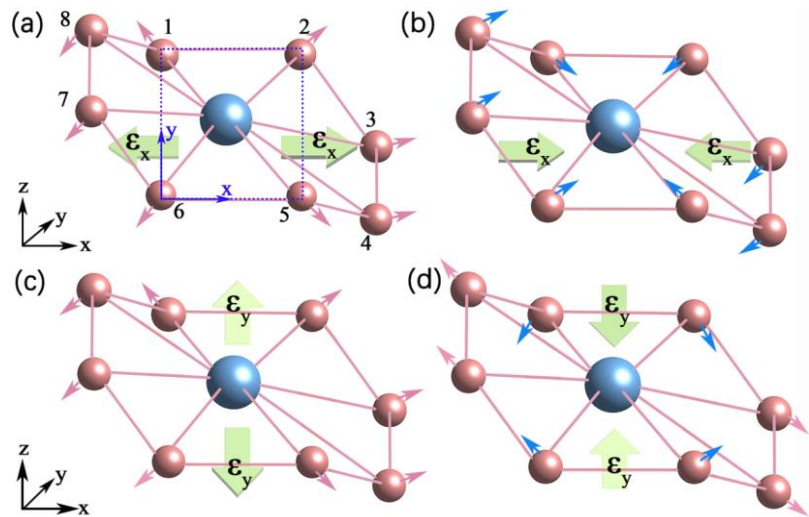

Fig. 3 The structural evolution of the Rh©B$_8$ wheel under uniaxial tensile/compressive strain along (a-b) *x* and (c-d) *y* direction. The green arrows represent strain directions. Solid pink/azure arrows display the atoms move outward/inward. Blue dashed frame represents the *x-y* plane.

Next, we consider the strains applied along the y direction (Fig. 2b). Under a tensile strain (Fig. 3c), the four in-plane B are stretched away from the center of the wheel. Due to the bonding between B sites, the four out-of-plane B move towards the z=0 plane, i.e., the thickness of the sheet naturally shrinks due to the in-plane stretch. Meanwhile, these out-of-plane B also have strong bonding with the central Rh, which maintains a more or less unchanged bond length ~2.12 Å (between Rh and B 3, 4, 7, 8) in this process. This leads to the observed expansion along x. Note that the reduction of thickness in this case is remarkably high, e.g. it reaches more than 30% under +8% strain along y. This explains the observed sizable negative PR. Finally, consider the compressive strain along y. The in-plane B 1, 2, 5 and 6 are moved inward under strain (see Fig. 3d). For the out-of-plane B 3, 4, 7, and 8, under the joint forces from the in-plane B and Rh, their natural choices are (1) move further away from the z=0 plane and (2) move outward along x. From our above discussion, the movement (1) may lead to a shrink along x, but it turns out that this tendency is subdominant here and is surpassed by the movement (2). This results in the observed positive PR for compressive strains along y. From the above analysis, it is clear that the unusual half-auxetic effect in RhB$_4$ is intimately connected with the special hypercoordinated Rh©B$_8$ wheel structure.

In the Supplemental Material, we further plot the evolutions of some key structural parameters under strain, which support our above analysis (Fig. S5-S6)[28].

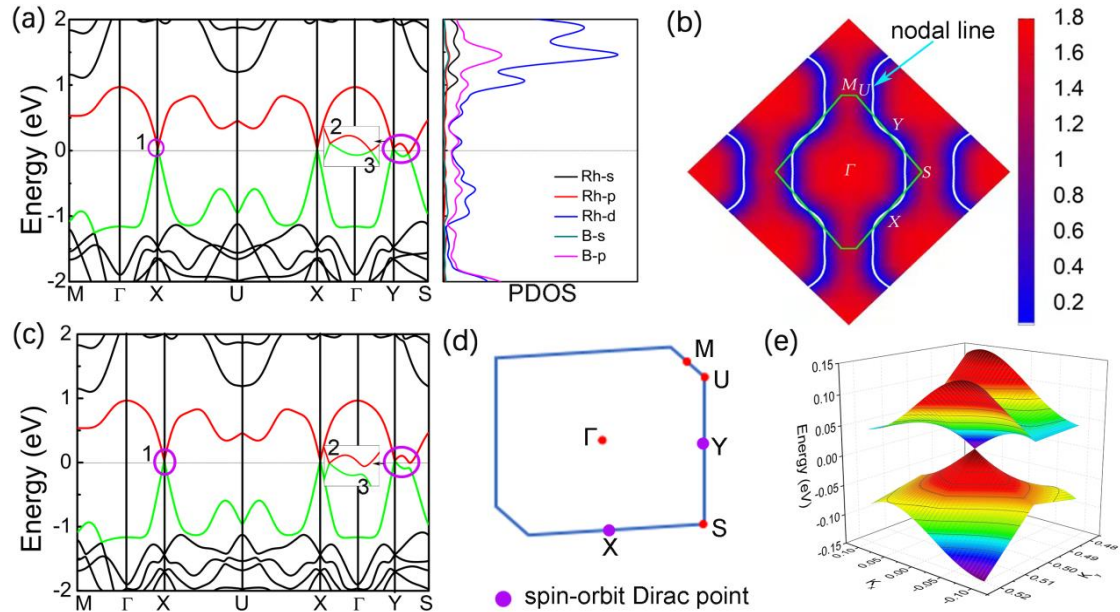

Fig.4 (a) Band structure and projected density of states (PDOS) of RhB$_4$ monolayer without SOC. (b) Shape of the double nodal lines obtained from the DFT calculation. The color map indicates the local gap between the two crossing bands. (c) Band structure of RhB$_4$ when SOC was considered. (d) The Brillouin zone for RhB$_4$. (e) The 3D band dispersion for the spin-orbit Dirac cone at Y point.

In the following, we turn to the electronic property of monolayer RhB$_4$. We will show that the corrugated structure of the material possesses nonsymmorphic space group symmetries, which enforce a 2D topological semimetal state.

The crystal structure of monolayer RhB$_4$ belongs to the *Cmma* space group. This group contains the following important elements: the inversion *P*, the glide mirror $\widetilde{M}_z \equiv \{M_z | \frac{1}{2}\frac{1}{2}0\}$, and the time reversal symmetry *T* is also preserved. Note that the glide mirror $\widetilde{M}_z$ is a nonsymmorphic symmetry, which involves a half lattice translation along both *a* and *b* directions. The presence of $\widetilde{M}_z$ is due to the corrugation in the wheel structure, because if the wheel were completely flat, $\widetilde{M}_z$ would reduce to a conventional mirror $M_z$ without the glide. In the following, we will see that the nonsymmorphic character of the symmetry is critical to the electronic state in the presence of SOC.

Let's first consider the band structure of monolayer RhB$_4$ without SOC. The result is presented in Fig. 4a. It can be clearly seen that the system displays a semimetallic character, with suppressed density of states (DOS) at the Fermi level. From the projections onto atomic orbitals, we find that the low-energy states are mainly from the Rh 4*d* orbitals (Fig. S7-S8)[28]. Note that despite possible correlation effects from Rh 4*d* orbitals, the system is found to be non-magnetic, as confirmed by our DFT+*U* calculations. Importantly, around the Fermi level, there are only two bands (four bands when counting spin), which are isolated from the other high energy bands. The two low-energy bands form several crossing points close to the Fermi level, as shown in Fig. 4a. In fact, these crossing points are not isolated. By carefully scanning the band structure in the Brillouin zone, we find that the points actually belong to two parallel nodal lines as shown in Fig. 4b, which is rarely seen in the previous studies. Here, each line is protected by the glide mirror symmetry $\widetilde{M}_z$, as the two crossing bands have opposite $\widetilde{M}_z$ eigenvalues. And the two lines are connected by the *P* or *T* symmetry.

When SOC is included, most points on the nodal lines will be gapped out. For example, a gap opening of 78.8 meV for Point 3 can be observed. However, one notes that Points 1 and 2 are still preserved under SOC (Fig. 4c). Here, Points 1 and 2 are fourfold degenerate Dirac points (sine each band here is doubly degenerate due to the *PT* symmetry) located at X and Y points of the Brillouin zone. Such Dirac points are quite unusual in 2D, and they should be contrasted with the Dirac points discussed in graphene. In graphene, the points are only stable in the absence of SOC and will be destroyed by SOC [39]. The points here are in fact realized only under SOC, and they are called the spin-orbit Dirac points [40,41].

The existence of the spin-orbit Dirac points is enforced by the three symmetries *P*, *T*, and $\widetilde{M}_z$. Note that each Bloch state can be chosen as an eigenstate of $\widetilde{M}_z$. The eigenvalue of $\widetilde{M}_z$ is given by $g_z = \pm i e^{-i(k_1+k_2)/2}$, since $\widetilde{M}_z^2 = T_{110}\bar{E} = -e^{-i(k_1+k_2)}$ where $T_{110}$ denotes the translation by one unit cell along both $a$ and $b$ directions, and $\bar{E} = -1$ is from the $2\pi$ rotation on spin. Here, we use $(k_1, k_2)$ to represent a wave vector with respect to the reciprocal lattice axis and $(k_x, k_y)$ to denote the Cartesian coordinate. To prove that the spin-orbit Dirac points are enforced at X and Y, it is important to notice two facts. First, the inversion does not commute with the glide mirror, instead, one finds

$$\widetilde{M}_z P = T_{110} P \widetilde{M}_z = e^{-i(k_1+k_2)} P \widetilde{M}_z$$

Thus, for an eigenstate $|g_z\rangle$ of $\widetilde{M}_z$ with eigenvalue $g_z$, the following relation holds:

$$\widetilde{M}_z PT|g_z\rangle = -e^{-i(k_1+k_2)}(PT)\widetilde{M}_z|g_z\rangle = -g_z(PT)|g_z\rangle,$$

which means the Kramers pair $|g_z\rangle$ and $(PT)|g_z\rangle$ are degenerate with opposite $\widetilde{M}_z$ eigenvalues at each $k$ point. Second, since the points X $(\pi, 0)$ and Y $(0, \pi)$ are both time-reversal-invariant-momentum (TRIM) point, any state $|g_z\rangle$ at X or Y must has another degenerate Kramers partner $T|g_z\rangle$ with the same eigenvalue $g_z$ (because $g_z = \pm 1$ is real at X/Y). The above two points ensure fourfold degeneracy at X and Y, with the linearly independent states $\{|g_z\rangle, T|g_z\rangle, (PT)|g_z\rangle, P|g_z\rangle\}$. Away from X and Y, $T$ is no longer preserved, and the fourfold degeneracy will generally split into two doubly degenerate bands. Hence, this leads to a fourfold degenerate crossing points at X and Y, which are the 2D spin-orbit Dirac points.

We can further derive a $k \cdot p$ Hamiltonian to characterize these spin-orbit Dirac points. Based on the quartet basis $\{|1\rangle, T|1\rangle, (PT)|1\rangle, P|1\rangle\}$ with $g_z = 1$, one can write down the matrix representations of symmetry operations as $\widetilde{M}_z = \sigma_z \otimes \sigma_0$, $P = \sigma_x \otimes \sigma_x$, $T = i\sigma_z \otimes \sigma_y K$ where $K$ is the complex conjugation and $\sigma_i$ are the Pauli matrices. Constrained by these symmetries, the $k \cdot p$ model to the first order in $k$ can be derived as

$$H(\mathbf{k}) = (c_1 k_x + c_2 k_y)\sigma_z \otimes \sigma_x + (c_3 k_x + c_4 k_y)\sigma_0 \otimes \sigma_y + (c_5 k_x + c_6 k_y)\sigma_0 \otimes \sigma_z$$

where the wave vector $\mathbf{k} = (k_x, k_y)$ and the energy are measured from the Dirac point, and the $c_i$'s are real model parameters. This effective model confirms the existence of spin-orbit Dirac points with linear band splitting at X and Y. In addition, we have also constructed a minimal lattice model, which captures the spin-orbit Dirac points in monolayer $RhB_4$. The details can be found in the Supplemental Material (Fig. S9-S10)[28].

We have a few remarks before closing. First, previously, 2D spin-orbit Dirac points were only found in monolayer HfGeTe family materials [40] and α-Bismuthene [42]. However, in those materials, the low-energy bands are not clean with the presence of extraneous bands around the Dirac points. For α-Bismuthene, the Dirac points are also away from the Fermi level. In this sense, the monolayer $RhB_4$ here represent the first almost ideal 2D spin-orbit Dirac point semimetal. Its low-energy band structure is very clean, involving only the bands that crossing at the Dirac points. The points are close to the Fermi level within 0.025 eV Thus, monolayer RhB$_4$ is so far the best system to study the spin-orbit Dirac point semimetal state.

Second, 2D spin-orbit Dirac points typically are not associated with topological edge states [40-42]. Here, we have checked the edge spectrum for monolayer RhB$_4$ and do not find any topological edge states near the Fermi level.

Finally, it is clear from our analysis that the special structure with hyper-coordination is a key factor underlying the formation of spin-orbit Dirac points in monolayer RhB$_4$. We have checked the possibility to replace Rh with other elements in this structure. It is important that the element should have the right valence to dope electrons to the B network, meanwhile, the element should also have the right size to fit into the voids in the B network. For example, Ir and Co are in the same column as Rh in the periodic table. However, the structure of IrB$_4$ is found to be unstable. And CoB$_4$ turns out to be antiferromagnetic, which breaks the time reversal symmetry required for the spin-orbit Dirac point. By screening through all *3d* and *4d* elements, RhB$_4$ is found to be only suitable candidate with the target structure and with spin-orbit Dirac points.

## IV. CONCLISION

In conclusion, we have investigated the structural, mechanical and electronic properties of the monolayer RhB$_4$ by means of DFT computations. Our discovered half auxeticity in monolayer RhB$_4$, i.e., the sheet always expands in one lateral direction when strained in the other in-plane direction, regardless of the sign of the applied strain. Our analysis attributes this exotic behavior to the corrugated Rh©B$_8$ wheel structure. Together with the finding in monolayer PdB$_4$, our work corroborates the structure-property relationship for the unusual mechanical properties in these hypercoordinated 2D materials. Furthermore, we show that the corrugated Rh©B8 wheel structure leads to a nonsymmorphic symmetry that underlies the spin-orbit Dirac points in the electronic band structure. The clean band structure around Fermi level makes monolayer RhB$_4$ the first almost ideal 2D spin-orbit Dirac point semimetal. Our result reveals the significance of hypercoordinated structural motifs in physical properties and predicts a new system, the monolayer RhB$_4$, for studying novel mechanical properties and topological electronic states.

## ACKNOWLEDGMENTS

The authors thank D. L. Deng for valuable discussions. This work is supported by the National Natural Science Foundation of China (Grant No. 11847017 and 11904077), Science Foundation of Hebei Normal University (Grant No. L2019B09), financial support program from Hebei Province (Grant No. E2019050018), Alexander von Humboldt-Foundation, and Singapore MOE AcRF Tier 2 (MOE2019-T2-1-001).

# Supplemental Material

# Monolayer RhB$_4$: half-auxeticity and almost ideal spin-orbit Dirac point semimetal

Zhen Gao,[1] Qianqian Wang,[2] Weikang Wu,[2,3,*] Zhixue Tian,[1] Ying Liu,[1] Fengxian Ma,[1,*] Yalong Jiao,[1,4] Shengyuan A. Yang[2]

[1] College of Physics, Hebei Key Laboratory of Photophysics Research and Application, Hebei Normal University, Shijiazhuang 050024, China

[2] Research Laboratory for Quantum Materials, Singapore University of Technology and Design, Singapore 487372, Singapore

[3] Division of Physics and Applied Physics, School of Physical and Mathematical Sciences, Nanyang Technological University, Singapore 637371, Singapore

[4] Faculty for Chemistry and Food Chemistry, TU Dresden, Bergstraße 66c, Dresden 01069, Germany

*E-mail: weikang.wu@ntu.edu.sg, fengxianma@hebtu.edu.cn

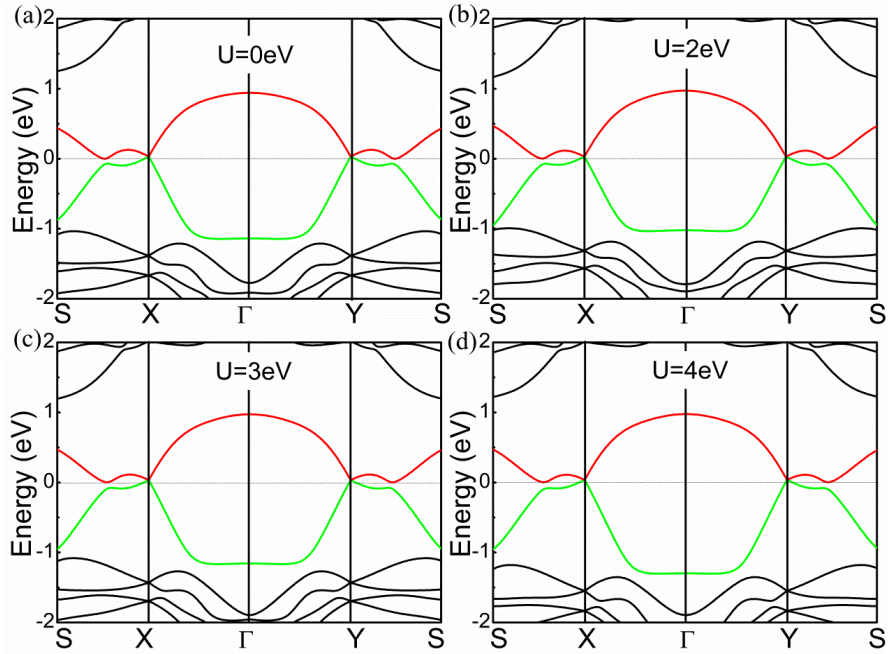

Fig. S1 (a-d) Band structures with SOC for RhB$_4$ monolayer when different on-site Hubbard U parameters (U = 0, 2, 3 ,4 eV) were applied.

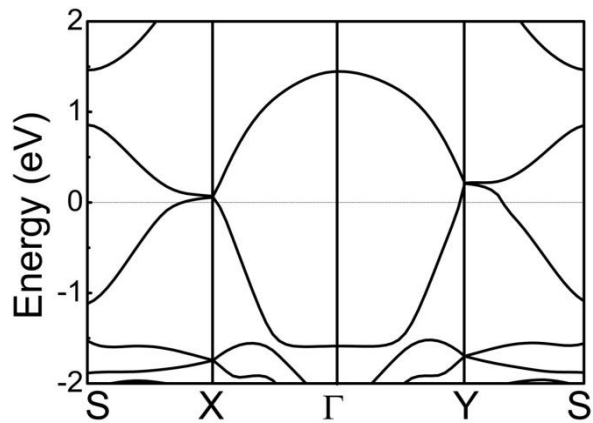

Fig. S2 Band structure of RhB$_4$ monolayer calculated by HSE+SOC method.

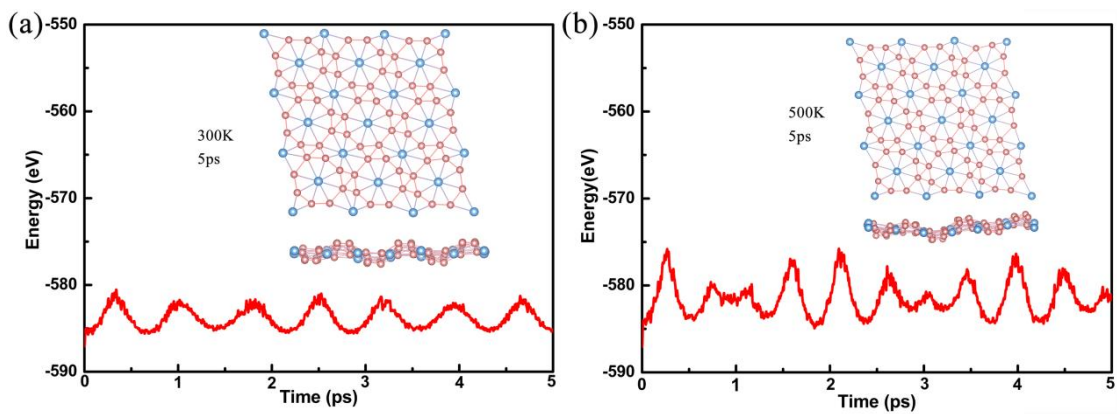

Fig. S3 Variation of the energy as a function of time for the RhB$_4$ monolayer at (a) 300K and

(b) 500K. The insets are the structure snapshot at the end of the AIMD simulations.

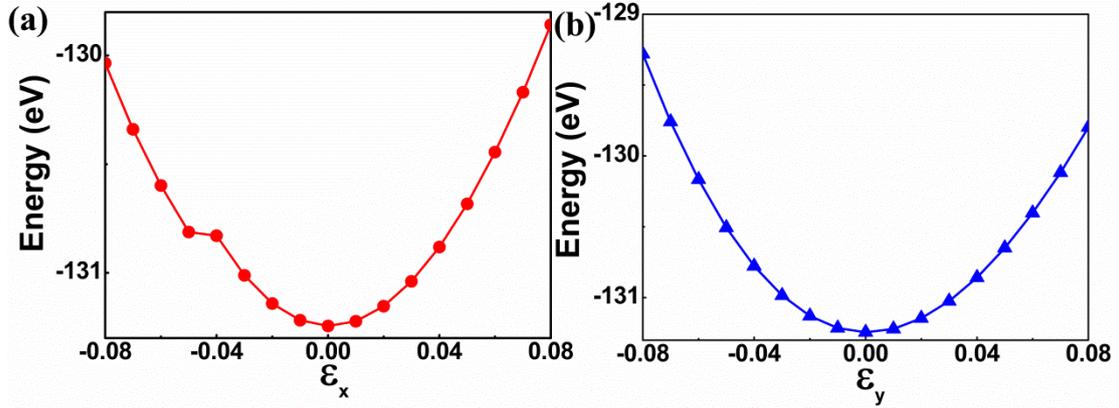

Fig. S4 (a-b) Energy-strain curves of the monolayer RhB$_4$.

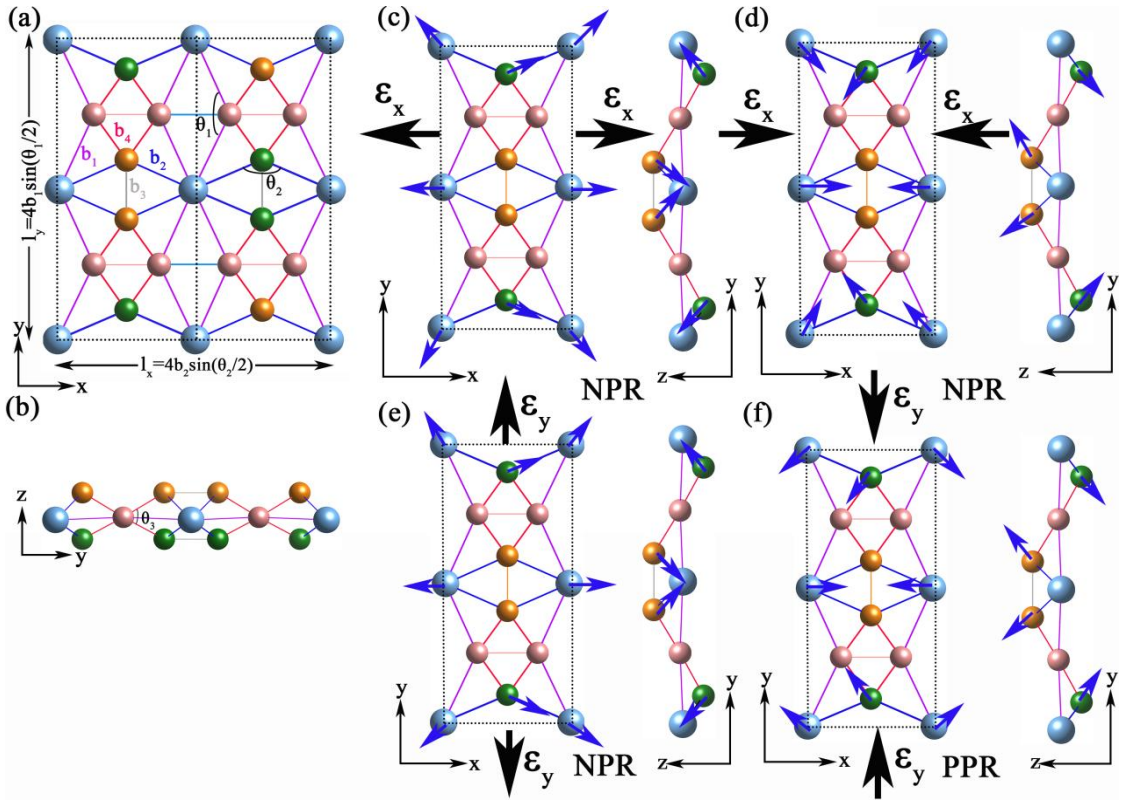

Fig. S5 (a-b) Top and side view of the unite cell of the strain-free RhB$_4$ monolayer. Boron atoms above, in, and below z=0 plane are represented by dark yellow, pink, and green, respectively. Two lattice vectors in the rectangular cell are directly related to the distance between Rh-Rh atoms. The structural evolution of RhB$_4$ under tensile/compressive strain is shown along (c-d) x- and (e-f) y-direction, respectively. For simplicity, the atom movements in the half of the unit cell is presented in (c-f).

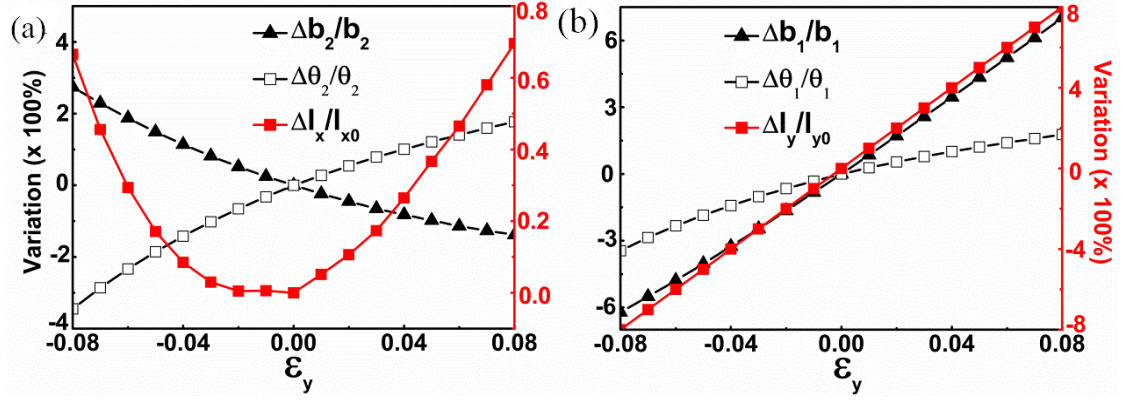

Fig. S6 Dependence of the strain along y direction (εy) of the bonding parameters $b_1$, $b_2$, $\theta_1$, and $\theta_2$ (for defifinitions, see Fig. S4a). The lattice vector length $l_x$, and $l_y$ were obtained directly through equation $l_x = 4b_2 \sin\left(\frac{\theta_2}{2}\right)$, and $l_y = 4b_1 \sin(\frac{\theta_1}{2})$, respectively.

The in-plane negative PR is mainly attributed to the change of lattice constants $l_x$ and $l_y$, which can be expressed as: $l_x = 4b_2 \sin\left(\frac{\theta_2}{2}\right)$, and $l_y = 4b_1 \sin(\frac{\theta_1}{2})$. When $\varepsilon_y$ is applied (Fig. S4a), during tension, it can be clearly seen the variation of $\theta_2$ is responsible for the negative PR as $l_x$ follows its increasing trend with the increased strain. During compression, however, the change of $b_2$ becomes much faster and the increased $b_2$ overwhelms the decreased $\theta_2$, leading to the increased $l_x$. In short, the half-auxecticity in RhB$_4$ layer is mainly due to the competition between the geometric parameters ($b_2$ and $\theta_2$).

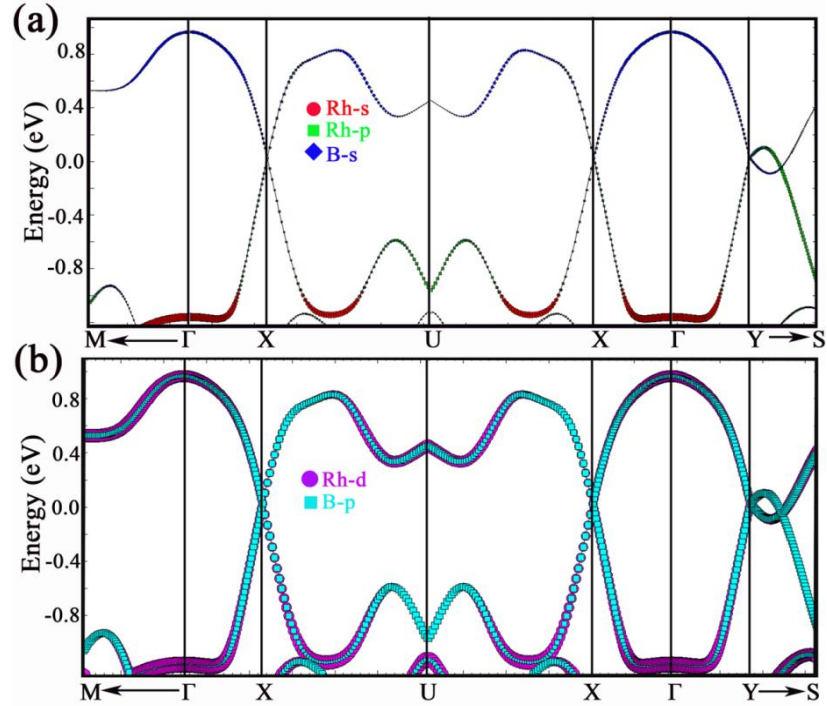

Fig. S7 (a-b) Orbital-resolved band structures for the RhB$_4$ monolayer.

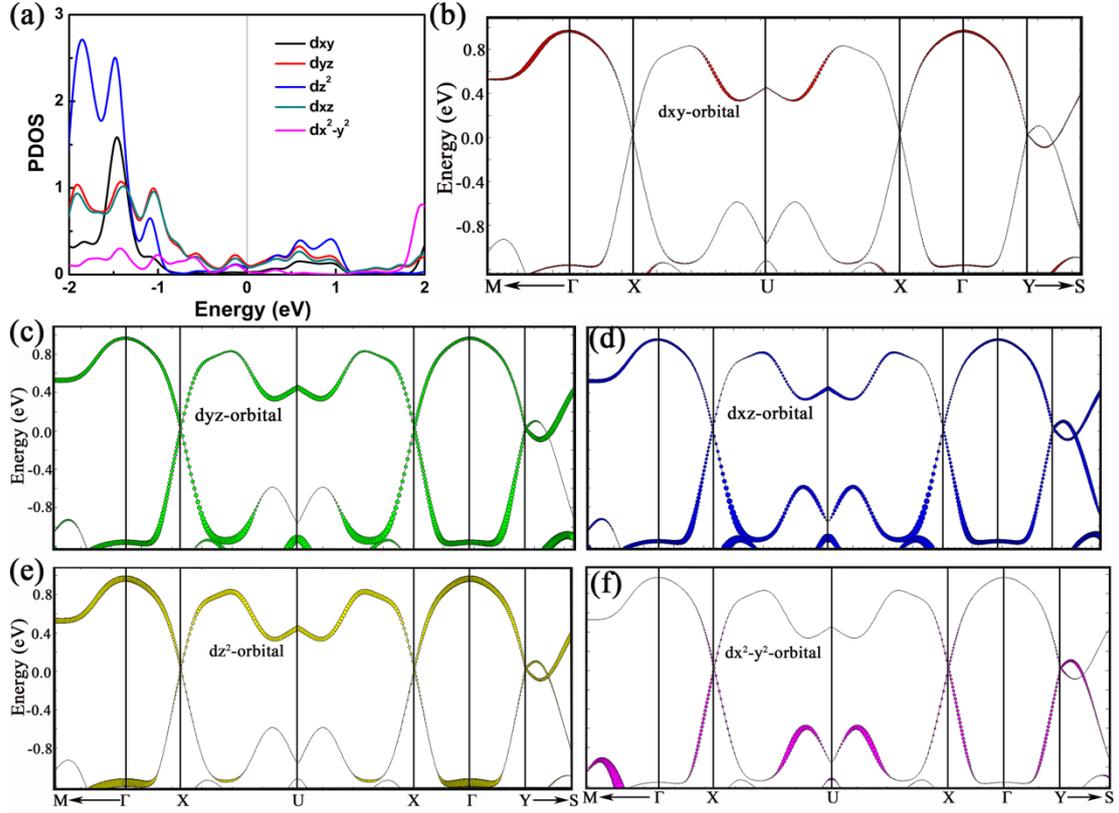

Fig. S8 (a) The PDOS for Rh atom in RhB$_4$ monolayer. (b-f) The corresponding orbital-resolved band structures.

**A minimal lattice model for the spin-orbit Dirac points in RhB$_4$ monolayer**

The crystal structure of monolayer RhB$_4$ belongs to the *Cmma* space group (No. 67). This group contains the following generators: the inversion $\mathcal{P}$, the glide mirror $\widetilde{M}_z \equiv \{M_z|\frac{1}{2}\frac{1}{2}0\}$, and the mirror $M_y$. Additionally, the time-reversal symmetry $\mathcal{T}$ is preserved.

Guided by the symmetry, we take a 2D lattice with lattice vectors:

$$\boldsymbol{a}_1 = \left(\frac{a}{2}, -\frac{b}{2}\right), \qquad \boldsymbol{a}_2 = \left(\frac{a}{2}, \frac{b}{2}\right),$$

and the corresponding reciprocal lattice is given by

$$\boldsymbol{b}_1 = 2\pi\left(\frac{1}{a}, -\frac{1}{b}\right), \qquad \boldsymbol{b}_2 = 2\pi\left(\frac{1}{a}, \frac{1}{b}\right),$$

where *a* and *b* are the lattice constants of the conventional cell (see Fig. R3). We take two active sites within one primitive cell at

$$A1: (0,0), \qquad A2: \left(\frac{1}{2}, \frac{1}{2}\right)$$

and basis at each site has two spin states. Therefore, we have four bands in total for this lattice model.

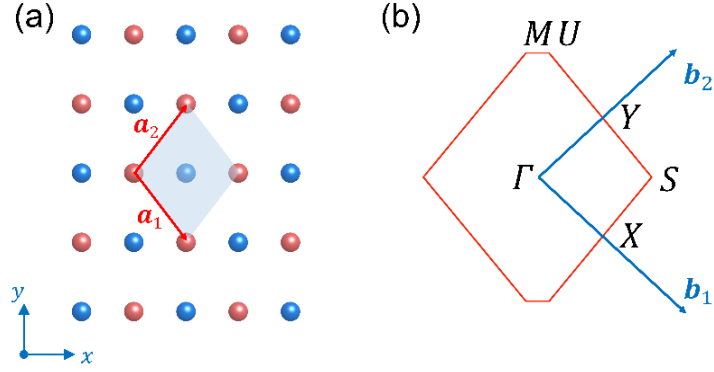

Fig. S9 The crystal structure for the lattice model and the corresponding Brillouin zone. The orange shaded area denotes the primitive cell.

In the basis ordered as $\{|A1,\uparrow\rangle, |A1,\downarrow\rangle, |A2,\uparrow\rangle, |A2,\downarrow\rangle\}$, the symmetry operators can be represented as
$\widetilde{M}_z = \sigma_x \otimes is_z$, $M_y = \sigma_0 \otimes is_y$  $\mathcal{P} = \sigma_0 \otimes s_0$, $\mathcal{T} = \sigma_0 \otimes is_y \mathcal{K}$
where the Pauli matrices $\sigma_i$ and $s_i$ act on the sublattice and the spin spaces, respectively, and $\mathcal{K}$ is the complex conjugation operation. Constrained by these symmetries, the symmetry-allowed Hamiltonian up to the second-neighbor hopping can be obtained as

$$H = \varepsilon_0 + (2t_1 \cos\frac{k_x a}{2} + 2t_2 \cos\frac{k_y b}{2}) \sigma_x \otimes s_0$$

$$+ 4t_3 \cos\frac{k_x a}{2} \cos\frac{k_y b}{2} \sigma_0 \otimes s_0 + (2t_1^{SO} \cos\frac{k_x a}{2} + 2t_2^{SO} \cos\frac{k_y b}{2}) \sigma_y \otimes s_y$$

where the coefficients $t_i$ and $t_i^{SO}$ are real-valued model parameters. A typical band structure for this model is plotted in Fig. R4 bellow, which successfully capture the spin-orbit Dirac points at X and Y.

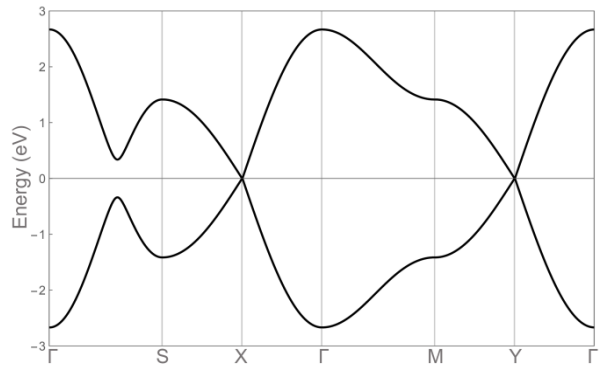

Fig. S10 A typical band structure for the lattice model. The parameters are taken as $\varepsilon_0 = 0, t_1 = 1.0, t_2 = 0.3,\ t_3 = 0,\ t_1^{SO} = -0.1$ and $t_2^{SO} = -0.2$.